# Comparison of 6 different approaches to outclass Top-k queries and Skyline queries

Martino Manzolini

Politecnico di Milano
Milan, Italy
martino.manzolini@mail.polimi.it

### Abstract

Top-k queries and skyline queries have well-explored limitations which recent research has tried to "complete" through new techniques. In this survey, after summarizing such limitations, we consider Restricted Skyline Queries, ORD and ORU approach, K-representative minimization queries, Skyline ordering queries, UTK queries approach and Skyrank that aim to overcome them. After introducing and comparing their main concepts, pros and cons, we briefly report the algorithms and compare some of the experimental data collected from the bibliography. To conclude the paper, we summarize the results presented with a short guide on how to select the best approach according to specific needs.

## 1. Introduction

In a fully connected reality, problems of simultaneous or conflicting aspects are common. For this reason, in the multi-objective optimization, concerning for example different attributes of the objects in a dataset, different approaches are used in all data-intensive contexts. For such data contexts, there are two main paradigms used to determine, for example in d-dimensional records, those which are the most interesting to the user based on dominance or ranking by utility. Different approaches used in different contexts, as written previously, consider different paradigms. Speaking, namely, of data mining and database systems, the most used approaches are the ranking queries (or top-k) approach and the Skyline approach. Both such approaches are based on the concept of the dominance paradigm which considers that a record dominates another if all its attributes are preferable: in other words, an object t it is said to dominate an object s if and only if t is no worse than s on all the attributes and strictly better on at least one. The top-k approach reduces the multi-objective problem to a single-objective problem by using a scoring function, based on the ranking by utility paradigm, used to express the relative importance of different attributes using parameters [13], and includes the records that are dominated by a maximum of $(k-1)$ other records. This linear type of scoring has been shown, by user studies, to model with efficacy the way humans ponder tradeoffs in real multi-objective decisions [13]. The skyline approach, instead, includes all the non-dominated objects

There are several drawbacks in the use of top-k and Skylines. Ranking queries, for example, depend a lot on the choice of weights considered in the scoring function, and this doesn't allow to have a global view of the dataset. Such lack of overall view of the dataset can be somehow fulfilled with the repetition the query several times with different choices of weights, perhaps by including in the scoring function other elements to consider, such as the diversity of the result set. The trouble, in this case, is that such elements also have problems of their own that can affect the final result. Also, still talking about specifying the "correct" weights of the scoring function, other problems can arise since a small change in the weights can heavily alter the top-$k$ result. The vector containing weights can be expressed by either direct input by the user or somehow mined through automatic techniques (e.g., online behavior and review mining, pairwise comparisons of example records or other learning methods [16]). The drawback stands because we can't expect a user to reasonably be able to quantify with total precision the various attributes and, on the other hand, the learning methods only output estimation of the mined weights.

Skyline queries provide a good overview of potentially interesting tuples and are effective in identifying interesting points in multidimensional data sets, but they also have known issues and weaknesses, such as they may contain too many objects. The result cardinality can be huge, in particular when the dimensionality is high, or the data is anticorrelated. Attempts have been made to resolve the problem formulating four categories of approaches. In pointwise ranking category [17][18], all d-dimensional points are ordered according to specific scoring functions and only the top points are returned, but this has the problem of reflecting the multiple criteria in the scoring functions. In subspace reference category [17][19], where all subspaces of the space are considered, the result is made from the points favored in the subspaces but with high computation costs. In the set-wide maximization category [20], a subset of the skyline is selected so that a collective quantity-based objective is maximized, having the issue though that requires a hard computational complexity resulting in producing only approximate result. In approximate selection category [21] instead, points are compared with a set threshold to have more points being identified as dominated but not being able to completely control the result cardinality. Everything is relative anyway, because also small skylines can cause not irrelevant issues, especially when the result requires a large number of objects, for example in the case of online shopping where the selection is made through filters. For this reason, given a fixed cardinality, a user may find that such a set does not offer enough alternatives and may instead wish a larger one, there is the need to formulate an approach to supports skyline querying for arbitrary cardinality k.

In this paper we recap 6 different new approaches, thoroughly studied studied in the references below, that aims to overcome the drawbacks just explained before through different solutions. Both the main characteristics and algorithms are compared, as well as the experimental results, in order to give, in the final section of this document, advice to the reader on which approach to choose.

## 2. Definitions and general comparison of the 6 new approaches proposed

Keeping in mind that ranking queries are less general and usually cheaper to compute than skylines, though widely used in data management systems, recently there have been studies trying to merge the previously mentioned approaches, aiming to correct their respective drawbacks. In this section of the document we are going to define six different approaches and compare their general advantages and drawbacks. In order to continue to define the differences and similarities more "in depth" between the previously mentioned approaches, we first have to introduce some fundamental concepts, taken from the sources consulted.

### 2.1. R-skylines approach

As a mix of skyline approach and top-k , there is the restricted skyline (R-skyline) queries [3]. R-skylines also consider the different importance that different attributes might have. Arbitrary families of scoring functions can be considered, and this introduces to a new concept of dominance called F-dominance, in which a tuple t F-dominates tuple s when t is always better, or equal, than s according to all the scoring functions in F. To use R-skyline we must consider two operators: ND, characterizing the set of non-F-dominated tuples, and PO, which represents to the tuples that are potentially optimal. When F is the family of all the monotone scoring functions then ND and PO are both equivalent to the skyline. On the other hand, if subsets thereof are considered, the behavior of one operator becomes different than the one of the other. R-skylines, in fact, consider all the practically relevant approaches to multi-objective optimization in a single framework and also allow the study of other scenarios of practical interest. In [3], referring to R-Skyline operators, is explained how their behavior is the same as Skyline, but applied to a limited set of monotone scoring functions $\mathcal{F} \subseteq \mathcal{M}$, where F is this set of monotone scoring functions and M the set of all monotone scoring functions. Intuitively, F is "rich enough" if, i.e. there is at least a function in F associating two different scores to two different tuples. There are also a few exceptions to this but are all cases captured by the framework. Then, by recalling the concept of F-Dominance resumed in the previous paragraphs we can introduce the R-skyline operators: "non-dominated restricted skyline", which consists of the set of non-F-dominated tuples in r, and "optimal restricted skyline", which returns the tuples that are best according to some scoring function in F.

## 2.2. ORD / ORU approach

Other studies of the previous kind try, by using other two operators, <u>ORD and ORU</u> [6], to satisfy the requirements of being output-size specified (OSS) [13], being personalized and having a relaxed preference input, aiming to blend together the main points of the previously mentioned dominance paradigm and ranking by utility paradigm. The acronym ORD points out its <u>O</u>SS property, <u>r</u>elaxed input, and <u>s</u>tronger dominance-oriented flavor. ORU abbreviation instead stresses the first two same characteristics of ORD but, by changing the last letter, specifies how it follows more closely the ranking by <u>u</u>tility paradigm. In [6], the operator ORD is explained, considering the seed vector w that contains the user preferences expressed by $d$ per-attribute weights $wi$, and the required output size $m$, to report the records that are $\rho$-dominated by fewer than $k$ others. The $\rho$ it's the stopping radius within which also other order-sensitive top-$k$ results are reported more than m exact records and is automatically determined by the proposed framework depending on the chosen $m$. The operator ORU, which instead follows more the ranking by utility paradigm, given w and m, reports the records belonging to the top-$k$ result no less than one preference vector within $\rho$ from w. ORU algorithm also can report the specific top-$k$ result for any vector inside the radius $\rho$ from w and. determine the most stable or the most representative top-$k$ results close to w. ORD/ORU approach, by trying to overcome the drawbacks of Skyline and Top-k approaches, proposes a new way that aims to achieve personalization, controllable output size and flexibility in preference specification for practical decision support.

## 2.3. K-regret approach

Another approach to deal with multi-criteria decision is the <u>k-representative regret minimization query</u> (k-regret [4]) that, just like top-k, assumes that users have some utility or scoring functions despite it never requires for the users to provide such functions. Similarly to skyline, it produces the result of filtering a set of relevant points from a potentially large database, according to the users' criteria. Despite that, it never engulfs the users with outputting too many tuples. Specifically, considering any number k and any class of utility functions, the k-regret query is able to output, from the database, k tuples trying to minimize the maximum regret ratio. Given a k-representative regret minimization operator (k-regret), the regret ratio of a user can be defined to be 1 minus the happiness of the user's utility function for the k tuples. So, the objective of such k-regret operator is to find k tuples to minimize the maximum regret ratio. Starting from a set of n d-dimensional points D, from the database and an output integer k, the goal is to find a set of tuples S of size at most k, that minimizes the maximum regret ratio. Knowing that a utility function is a mapping f : Rd + → R+, that the gain of a user is the maximum utility derived a subset of points and that the regret ratio, which measure the "satisfaction" of the user about such points spanning from 0 to 1, we use the maximum regret ratio to define the approach. Having no information of any distributional information on the utility functions, the focus is to keep the maximum regret ratio low, assuming that k is at least d and focusing on the class of linear utility functions as defined in [4].

## 2.4. UTK queries approach

Another approach is the underline{uncertain top-k query (UTK)} [23] that, in a first version of the problem ($UTK_1$), given uncertain preferences which are an approximate description of the weight values, reports all options that may belong to the top-k set . A second version of the problem ($UTK_2$) additionally reports the exact top-k set for each of the possible weight settings. Since the the assumption/requirement that the exact weight vector is known, in top-k queries, is hardly realistic, or anyway inherently inaccurate if specified by the user, UTK instead expands the weight vector into a region and report all possible top-k sets to the user in order to offer the possibility to user to explore other similar alternatives. Thus, practically means to provide the user with top-k results for similar preference profiles. In [23], through an example concerning hotels and some of their attributes, is explained how, given as UTK input the value k (highest-scoring records form the output of the top-k query ) and the region of interest R, the approximated or expanded user preferences are represented with convex polygon in the preference domain. Two different UTK are distinguished: the first reports all the hotels that may belong to the top-2 set, considering that the result must be minimal, meanwhile the second provides more detail reporting the precise top-2 set for any possible positioning of the vector w of weights in R. Is relevant to state that nothing is probabilistic in UTK and that the records' scores are correlated and change together as w is freely positioned inside R.

## 2.5. Skyline ordering approach

Further methods base their definition on the conception of size constrained skyline queries that use an arbitrary size constraint k, taken as parameter, and find k searched points from a dataset of d-dimensions. Starting from the assumption that when a specific k is chosen, the original skyline of the data set is unknown, as consequence it is also unknown if the parameter k is larger, equal, or smaller than the original skyline. Said that, underline{skyline ordering} [1] is a new approach that works with constraints having arbitrary size rather than with the classical skyline query. This approach, by adopting a skyline-based partitioning of the data set, allows partitions to be ordered among themselves and also reserves space for using different set-wide maximization techniques inside those partitions.

As written in [1], the skyline order of a set of d-dimensional points P is a sequence $S = <S_1, S_2, ... S_n>$ where each $S_i$ is a skyline subset in the skyline order or a skyline order subset, and n is the skyline order length, i.e., the number of subsets in S. The definitions iteratively remove the current skyline starting with P until all points in P belong to a skyline subset, as a result of which n skyline subsets are obtained.

The objective of Skyline ordering is to overcome the drawbacks of pointwise ranking and set-wide maximization by merging them both into a homogeneous framework. We have to start by considering that, in skyline order, no point can dominate any other point in the same or previous partition, and that any point in a partition, except in the first one, must be dominated by some points in the previous one. In this approach, indeed, we start from the first partition and then we keep outputting partitions until, as minimum, k points have been output. This way is possible to compute customized size constraints on skyline in an efficient way within a flexible framework, even though this may involve several consecutive partitions.

## 2.6. Skyrank approach

As last alternative to skyline queries presented in this paper is underline{Skyrank} [2], a framework for ranking the skyline points when we lack a user-defined preference function, using a limited subset of the most interesting points of the skyline set. This is a method that compares and ranks the skyline points of a dataset using a weighted directed graph called skyline graph, which aims to represent the dominance relations between the skyline points for different subspaces. Skyrank applies well-studied authority-based ranking algorithms on the skyline graph and highlights how important it is for a skyline point to exploit the subspace dominance relationships. By applying a link-based ranking approach to the skyline graph automatically means that dominated points transfer their superiority to the dominating points since they lose their superior status in the subspaces. Furthermore, it is possible to extend Skyrank to handle top-k preference skyline queries, when the user is able to specify certain preferences. In [2] is described how, in Skyrank, the exploration of the subspace dominance relationships between pairs of skyline points is necessary in order to discover the importance of skyline points. The interestingness of a skyline point based on the dominance relationships in difference subspaces can be intuitively expressed by saying that a skyline point is more interesting if it dominates several other skyline points in many subspaces, and these points also dominate other points in other subspaces. Such interestingness is recursive, and the scoring values are calculated by an iterative algorithm. In Skyrank, all dominance relationships are mapped into a weighted directed graph, called skyline graph. Such graph will be explained more in detailed in the section dedicated to the algorithms

## 2.7. General comparisons between the six approaches

From these definitions are evident some differences: meanwhile Skyline Ordering Approach proceeds by removing skylines starting with a certain set of points, with an iterative process, until all such points are part of a skyline subset, also Skyrank works through iterative analysis but by comparing pairs of skyline points arbitrarily, in order to find their "importance" and lately the final set of points composing the skyline through dominance. In R-Skyline instead, as mentioned before, such "importance" is determined through the concept of F-Dominance by computing ND or by computing the F-dominance regions of tuples and discard those that belong to at least one region. Meanwhile PO, for any set F, can be computed from ND by retaining only the tuples that aren't F-dominated by any virtual tuple found by combining other tuples in ND. In ORD/ORU differently, the choice of importance is based on a tradeoff between the paradigms of dominance-based and ranking by utility, using these two operators, trying to include the concepts of being OSS, being personalized and having a relaxed preference input. In K-regret approach instead the interest seems to be more focused on the "happiness" of the user, using the regret operator. In this approach the important is the attempt to finding k tuples to minimize the maximum regret ratio, which is indeed equal, as explained previously, 1 - such happiness of the user's utility function for those k tuples. Something related happens also in UTK approach where, similarly to K-regret approach, the focus is to provide a user-centric and practical design but not through a regret operator but by computing weight region to consider uncertain preferences.

The ranking of the skyline points in Skyrank does not get modified when points that do not have an impact on the skyline set are added to the dataset. This allows it to master the attribute of robustness to updates, since only in the case the skyline set changes the ranking of the points requires to be updated. This is particularly important for that approach that is built to consider vast datasets and indeed a knowledge about the entire dataset is not needed, but only about the skyline points. Similarly, also in the case of the skyline ordering, the original skyline of the data set is unknown and for this reason it is also unknown the constraint k size, making it good for the unpredictability. On the contrary of Skyline Ordering, for using ORD/ORU no precomputation is needed, just a general-purpose spatial index on $D$, so, updates in $D$ affect only such index. This technique also enables the integration of common predicates into our framework, but since multi-objective querying loses its meaning in high dimensions, the focus is on low-dimensional settings. In R-Skyline approach instead, the advantage focuses trying to find a tradeoff between attributes, similarly at [9], such as the vastity in the dataset and a less general but cheaper analysis, rather mastering one only. In k-regret approach, similarly to Skyrank, the size of the dataset is independent from regret ratio guarantee, and so also the happiness guarantee, since even if some values in some attributes are multiplied with a constant, k-regret will output the same set of tuples anyway. Another trait that the two approaches have in common is the stability which is a certain insensitiveness to adding or deleting junk points. UTK approach instead, has a similarity with Skyrank in the ability of being efficient while considering voluminous datasets. An evident difference with Skyline Ordering is that the considered uncertainty in this case concerns the size of the dataset, meanwhile in the case of UTK it's about the uncertainty of the weight values.

There are similarities in the dimensionalities of the dataset considered for analyzing all the approaches presented. In Skyline ordering there are no in-place requirements, and the algorithms are not limited to two dimensions datasets. Also Skyrank approach, whom analyses the results by structuring a graph generated through an algorithm explained in the further paragraph, which aims to show a 2 dimensional projection of the skyline points, is based on a d-dimensional dataset. Similarly, experiments are conducted on d-dimensional dataset in R-Skyline, and the other two approaches. The cardinality, instead, in all the graphs presented in the papers in the bibliography, varies abundantly from experiment to experiment.

## 3.  Comparison between the 6 different approaches algorithms

In order to continue with comparisons and analysis, it is necessary to first briefly introduce a short explanation on how the algorithms work.

### 3.1. Skyrank algorithms

To understand more about the Skyrank algorithm is important to touch on the concept of Skyline graph. For the sake of simplicity previously mentioned, is omitted its construction algorithm which can be studied by consulting [2]. Considering a data space D of a randomly generated dataset S, and a projection of the skyline points of S on subspace U as shown in [2], the domination of the skyline points in the subspace can be represented with a skyline graph. In such graph vertices of different colors are connected by arrows, ingoing or outgoing the nodes, and those vertices usually colored of grey are the skyline points. Inlinks and outlinks represent the domination relations and the number of outlinks of a vertex is the same of the points that dominate it.

Knowing how the subpart of a skyline graph is composed, is intuitive how the creation of such graph is relevant from a computational point of view. Once the graph is computed, the framework Skyrank can be adapted to user-defined preferences, without recompute the skyline graph, allow a certain flexibility and spare of computational resources. Calling the maximum number of skyline points in each subspace is |SKY|, the upper limit of dominance test is $\ell(d) * |SKY|$, where $\ell(d)$ is the number of different parent-child relations. It is more less uncommon that all subspaces have |SKY| skyline points, making this is the worse-case scenario where all skyline points have the same value for all dimensions. Even if users may prefer some particular subsets, for example while looking for a hotel may prefer a cheap one close to the beach but also with high star rating, Skyrank can easily adapt to such preferences, without requiring defining manually specific weightings. Therefore, the user needs only to specify a set of m preferred subspaces PRF={U1, …, Um}, with Ui ⊂ D. Differently than other well-known algorithms used for ranking, findable in the bibliography, the Skyrank scores are indeed biased in favor of the skyline points of the preferred subspaces.

### 3.2. Skyline Ordering algorithms

Speaking about the Skyline ordering approaches, there are 5 algorithms in total that has to be considered, described in [1]. Once again, for the sake of ease in the exposition of the main purpose of this paper, the focus will be put on the comparison of pro and cons and not on the formalities of the algorithms. We briefly analyze the proposed skyline order computation algorithms. Unlikely the Skyrank approach, Skyline ordering algorithms do not use a skyline graph as input or to represent the dominance. The first 3 algorithms are respectively SkylineOrderScan, AdjustSkyOrd and SkylineOrderPresort [1].

In SkylineOrderScan algorithm, the best case is that each point p, from an input data set P, forms a new first skyline order subset in S, without invoking the procedure AdjustSkyOrd. In this case the total number of dominance comparisons, processing the whole data set P, is N - 1. The worst case, involving cascading adjustments caused by a call of AdjustSkyOrd, leads to the same worst-case complexity for both of the algorithms of O($N^3$). For the SkyOrdPreSort algorithm, respectively without and with binary search, the best case i.e., N-1 and $\log_2$(N!) dominance comparisons. Its worst case, if each point p from P is compared with every point in every skyline subset, leads to a total of N(N-1)/2 dominance comparisons for both.

Considering the last two algorithms that process size constrained skyline query using the concept of skyline order, which are called SCSQuerySkyOrdPre and SCSQuerySkyOrd, best case and worst case will be reported in the paragraph concerning the experimental evaluations.

### 3.3. R-Skyline algorithms

In R-Skyline approach, in order to compute ND and PO, there are different alternative algorithms.

Starting with the computation of ND, the first option is to sort the dataset beforehand to have a topological sort of F-dominance relations. The sorting function is a weighted sum where the weights the coordinates of a polytope [14], explained in [3], where if a tuple t precedes a tuples s in this order, s is not dominated by t. Another option regards the F-dominance test where either the dominance is checked by solving an LP problem or using the F-dominance region. Or it can be chosen if the computation of ND should be applied after computing Sky or if dominance and F-dominance tests should be integrated. There are mainly 2 varieties of algorithms: sorted and unsorted variants. Sorted algorithms scan the tuples sordidly and fill a current window ND of nondominated tuples among those that are in Sky(r) (skyline of the r, instance of the dataset), this way so that no tuple will be removed from ND thanks to sorting. We then enumerate sordidly all candidates F-dominated tuple s and compare it with all candidates F-dominant tuple t to see if s should be added to ND. Unsorted algorithms are similar but without sorting. In this case when a tuple s is added to ND, other tuples can be dominated by s, and so they need to be removed. For computing PO, we start from the tuples in ND, we discard any tuple t that is F-dominated by a convex combination of tuples in ND \ {t}. For fastness's sake, the set of candidate potentially optimal tuples is reduced by starting with a convex combination of only 2 tuples, by sortedly enumerating candidates in reverse order and by checking the existence of a convex combination of the first tuples as they are the best in the ordering and the ones that probably will F-dominate others. To finish, all the remaining tuples are compared with a convex combination of all the other tuples still in PO. The complexity of the computation of ND, which basis computation concepts can be thorough in [10], and even more detailly in [22], generally is less in the sorted algorithm but can vary depending which phase of the elaboration is considered (first or second). More specific data, not inserted in this document otherwise more formal notation would have needed to be explained, can be consulted in [3]. In [3] not much data is instead reported, especially in the experimental section, to explain in detail the complexity of the computation of PO, but in [22] is stated an intrinsic higher complexity of computing PO with respect to ND.

### 3.4. ORD/ORU algorithms

In ORD/ORU, the algorithms for processing the two operators are several, here will be resumed only those that are more efficient for ease's sake and practical evaluation. Starting with ORD is necessary to avoid computing, in the start of the process, the entire $k$-skyband which may many more records than the $m$ required. To do this, a progressive $k$-skyband retrieval is invoked to fetches its members and place them into a candidate set in decreasing score order for w, so that when a new candidate r$i$ is fetched, its inflection radius $\rho i$ can be immediately computed. The fetching stops when the set reaches size ($m$ + 1), then we eject the candidate with the widest inflection radius ri. Since the maximum inflection radius $\rho$-skyband includes at least the $m$ existing candidates, we switch from fetching $k$-skyband members to $\rho$-skyband ones because its retrieval becomes increasingly more selective, and when it cannot fetch any more records, the candidate set becomes the ORD result. ORU requires to produce an overestimate of $\rho$, to ensure a minimum output of size $m$, which can be the radius required so that ORU's output for $k$ = 1 includes $m$ records. Ideally, to derive such overestimate of $\rho$, the best would be to localize the upper hull computation close to w by exploiting the fact that the $\rho$-skyline is a superset of ORU's output for the same $\rho$ and $k = 1$ using an algorithm detailed in [6] and prompting it until the $\rho$-skyline includes $m$ records. Next, we compute their upper hull $Ltmp$. We use as overestimate of $\rho$ the final radius reported compute the $\rho$-skyband, through a $k$-skyband algorithm, then take its members into a candidate set $M$. The last part can be found in [6] but will be omitted since requires several paragraphs of explanation.

### 3.5.K-regret algorithms

About K-regret, in [4] is explained how to show the maximum regret ratio is used an algorithm called Cube algorithm which outputs a set S of size at most k. First, it outputs the point with maximum value in each coordinate, leaving the last coordinate, then it divides every dimension into t equal-sized intervals, excluding the last dimension. It continues partitioning the points into "buckets" based on the intervals in each dimension and outputting the point with the highest value, in the last coordinate, in each bucket. Since the points in the same bucket have similar gains, when it chooses the point with maximum value its gain is close to the one of other points in the same bucket. The regret is then bounded. Another algorithm used in k-regret approach is the Greedy algorithm and is based natural greedy heuristic. By following the framework of Ramer-Douglas-Peucker algorithm explained in [8, 11], it approximates curves and polygons which has been observed to work well both in 2 and higher dimensions, using the maximum regret ratio as a measure. Greedy starts by picking the point that maximizes the first coordinate and then, through different iterations, adds the the point p that every time contributes to the maximum regret ratio to the solution set S. The point with maximum value is kept with similar intention of the previous algorithm.

### 3.6.UTK queries algorithms

Starting from $UTK_1$ [23], to understand the r-skyband algorithm (RSA) is required the knowledge of r-dominance. On the contrary of traditional dominance, where the dominator is preferable to the dominee for any weight vector w, r-dominance is instead specific to weight vectors in the region of interest R. A record p1 r-dominates another record p0 when S(p1) >= S(p0) for any weight vector in R, and this is important concept since the only records that could be in the $UTK_1$ result must belong to the r-skyband of the dataset which, by definition, includes only those records that are r-dominated by fewer than k others. On the other hand the r-skyband can only be used as a filtering tool for $UTK_1$ processing, which makes it the first step of the algorithm. The computation of r-skyband is similar to k-skyband computation but using a different visiting order for R-tree nodes and records. A second step thorough in [23] shows how, since RSA considers candidates one by one in decreasing order of their r-dominance counts, r-dominance is also useful for a refinement of the computation of the result improving further the efficiency of the algorithm. For the $UTK_2$ problem instead, in [23] is explained the Joint Arrangement Algorithm (JAA) which partially overlaps the RSA concerning the filtering step, meanwhile the refinement is fundamentally different and the output of JAA is a single partitioning of R, where each partition is associated with the respective top-k set. The output of $UTK_2$ it differentiates from the output of $UTK_1$ since it is more explicit/detailed.

## Algorithms' comparison

Starting by considering Skyrank and Skyline ordering methods, a clear difference in the calculation of the best-worst case between the two approaches is that, meanwhile in Skyrank approach it depends also by the parent-child relations, in the Skyline order algorithms only depend on the cardinality N of the dataset. The dominance test's numbers, which define the complexity of the algorithm, can be variable between the Skyrank approach and the Skyline ordering ones. If we consider a fixed cardinality N, by using Skyline ordering (both algorithm without and with binary search) we generally have a less complexity than by using Skyrank, according to the data provided by [1] and [2] and reported before. In the case of R-Skyline, the complexity also depends on N but also on the number of constrains c. If, for the sake of comparison and generalization, we consider only the phase of the less performing algorithms of R-Skyline where the complexity depends only on N, we obtain a lesser value than in Skyline Ordering approaches. In ORD/ORU the complexity is once again determined by N and other parameters as well such as dimensionality like the others, parameter k and output size m. Unlikely Skyrank, Skyline ordering and R-Skyline, in the case of k-regret the complexity of the algorithms generally doesn't grow as a power of N but, only in the case of Greedy, complexity grows as a power of dimensionality and k which potentially makes it, if N is high but d is low, less computational complex than first three algorithms previously mentioned. In UTK, the complexity doesn't only depend by N, which determined inevitably the number of the candidates, but also by the dimensionality which appears as exponent during its computation.

## 4. Experimental evaluation comparison between the 6 different approaches

Considering the crucial differences between the way to operate of the approaches, the comparison between the evaluation of the algorithms of this section will be done only through the cardinality/dimensionality and time data. The data collected from the reference papers, on this matter, are the clearest and shortest to compare without the need of several other assumptions and considerations to add. Some papers offer real dataset, but some others offer only generated ones. When they are provided, I will use the real one, otherwise I have no choice but to consider the generated ones: considering both for every approach would double the length of this section. Another premise necessary to make explicit is that, in some approaches, the datasets are calculated both independent and anticorrelated, obtaining different results during the test for each. In the following comparisons I will try to report such results providing the range of data in between, instead of specific results for this approach. Finally it is important to state that, considering all the mentioned heterogeneity in datasets, hardware and scenarios, the results presented cannot be considered 100% accurate and so cannot be completely trusted.

Starting by comparing the different assumptions on which the tests taken from the bibliography are made, in the following section are collected such data divided by approach.

| | Skyline ordering approach | Skyrank approach | R-Skyline approach | ORD and ORU | K-representative regret minim. query | UTK queries approach |
|---|---|---|---|---|---|---|
| Hardware | 2.8 GHz processor 1 GB RAM Windows XP | 3 GHz processor 2 GB RAM Windows XP | 2.2 GHz processor 16 GB RAM Not specified | 3.60 GHz processor 32 GB RAM Not specified | 1.7 GHz processor Not specified Linux 2.6.9 | 3.4 GHz processor 16 GB RAM No specified |
| Datasets | NBA dataset of players from 1946 to 2004 | Generated through a method explained in [5] | NBA dataset of players from 2008 to 2015 | NBA dataset of players in 2018-2019 season | Dataset from Dr. Yufei Tao's homepage | NBA dataset [24] |
| Dimensionality | 3-5, 10,15, 20 | 2-7 | 2,4,6,8,10 | 2,3,4,5,6 | 2-10 | 2,3,4,5,6,7 |
| Cardinality | 100K – 1000K | 50K-150K points | 10K – 1M | 100K – 25.6M | 100-1M | 100K-1600K |

Starting with skyline order computation approaches (the basic scan approach, the improved approach with presorting, the one with binary search in addition to presorting, the algorithm with precomputed skyline order and the algorithm without precomputed skyline order) we try to get generic conclusion over the computation time cost.

### 4.1. Experiment on Skyline Ordering's algorithms

The effects of data set cardinality in Skyline ordering, as it increases, the basic scan approach deteriorates dramatically, the binary search approach only slowly incurs higher cost, and the presorting approach remains in between. The presorting approach performs better than the basic scan approach, much more than the improvement between the binary search and the presorting one. In general, if considering the "more advanced" algorithms of Skyline ordering, the cardinality affects relatively mildly the performances of the approach [1].
About precomputation-based approaches usually outperform their corresponding counterparts not having precomputation, this because precomputed skyline orders save query processing time. In this case the data provided by [1], considering both the algorithms, show a range of time between 0.05 and 5 s at the lowest cardinality and between 0.1 and 10 s at the highest cardinality, where the lowest points are represented by the precomputation based.

As the data dimensionality increases, instead, after 2 the binary search is no longer the best and after 10 becomes the worst, meanwhile the basic scan approach becomes comparable to the presort approach. For dimensionality 10-20, the very short skyline order lengths allow only very limited room for optimizations. In this case, after certain empiric thresholds of dimensionality, the performances worsen more using advanced algorithms [1].
About the algorithm with precomputed skyline order and the algorithm without precomputed skyline order, as the dimensionality grows, both deteriorate evidently. The count-based reduction incorporated by these approaches requires more time to check for dominance when dimensionality increases meanwhile if the dominating capability is estimated, the cost even decreases in some cases but grow again after the middle of the x-axis considered. Trying to draw an average from the data in [1], the range goes from 0.1-5 s at the lowest dimensionality to 1-500s at the highest dimensionality.

### 4.2. Experiment on Skyrank's algorithms

Referring to Skyrank approaches, the construction time of the skyline graph for our algorithm depends mainly on the dimensionality of the dataset and is only only slightly influenced by the cardinality because it affects mildly the skyline size. On the other hand, the increase of the dimensionality causes a rapid increase of the skyline points and of the number of subspaces that have to be examined. In the section of [2], where is shown evaluation of the SKYRANK algorithm, the cardinality is fixed to 100000 and is considered only the variation of the dimensionality. In this case the query response time, similarly but more mildly in respect to the construction time of the skyline graphs, grow more less proportionally with the increase of the dimensionality starting with 0.05 s and arriving at the maximum dimensionality with 1000s.

### 4.3. Experiment on R-Skyline algorithms

About R-Skyline approach, similarly to the previous approaches, we evaluate the efficiency of the algorithms for computing ND by measuring the execution time. In all the categories of algorithms of R-Skyline the time increase as the cardinality and dimensionality grow, and also for the number of constraints c. In [3], and with more details in [22], starting from a dataset with cardinality from 10K up to 100K tuples, at min N the values of time range from 0.1 -100s with uniform distribution and between 1 – 1000s with Anti-correlated distribution. Similarly, at 100K tuples, values range between 0.5-100s and 10-10000s. By varying the dataset from 10K up to 1M tuples, some sorted algorithms now behave more efficiently than others in the anti-correlated dataset but the main differences is that in this last type of dataset, time grows proportionally to the growth of N meanwhile in the uniform dataset the growth seems exponential. In the same order of before, here values are ~0s and 0.1-0.8 s at 10K, meanwhile at 1M the two ranges are ~5s and 50-1000s. In almost the same trend, at the variation of the dimensionality, with a minimum of 2 and a maximum of 10, the execution time has the following ranges in the uniform dataset: ~0.2s at d=2 and 2-10s at d=10, and the following ranged in the anti-correlated dataset: ~0.2 at d=2 and 30-100s at d=10. Also, in this case the growth of time corresponding to the growth of d, seems exponential in uniform dataset.

### 4.4. Experiment on ORD/ORU algorithms

In the same way, considering ORD/ORU I try to resume a qualitative evaluation on the graphs found in [6]. Starting from ORD the running time grows sub-linearly to the cardinality, spanning between a range of 0.1-10s at minimum cardinality of 100000, and 0.5-1000s at maximum cardinality of 25600000. The growth of dimensionality instead provokes a more sharply, reaching ranges between 0,01-10s at minimum cardinality of 2, until having 1-1000s range at maximum cardinality of 7. Considering ORU, the graphs from the bibliography shows how the general trend of growth of running time due to cardinality is similar to the one of ORD, meanwhile for the dimensionality tend to be even sharper. In this case, with a minimum cardinality of 100000 the range spans between 1-1000s and with maximum cardinality between 50-10000s. At minimum dimensionality of 2 we have a range 1-50s and at high dimensionalities we have a range between 10-10000s.

### 4.5. Experiment on K-Regret algorithms

From [4] is possible to discuss the running times of various algorithms of K-regret approach. It is important to consider that in the experiments are conducted computing skyline first before running the algorithms, as discussed in [12] and [7] since minimizing the maximum regret ratio takes way less time than skyline computation. The running time on anti-correlated data, doing an average between Cube and Greedy, for the minimum cardinality of 100 spans between 0.0001-0.1s and between 0.5-100000s for the maximum cardinality of 1M. The graphs show how the time grows basically proportionally with the growth of cardinality. Varying the dimensionality instead, at the minimum value of 2 correspond a time range between 0.0001-0.1s meanwhile for the maximum value of 10 correspond a time range between 0.01-100s. Similar experiments, even reaching higher dimensionality, are reported in [7]. In the case of the dimensionality instead, the growth of running time starts harshly but the sharpness of the curve decreases with the increasing of the dimensionality.

### 4.6. Experiment on UTK algorithms

In [23] experiments are done on both the algorithms of UTK, and from the few graphs presented it is possible to observe how, both in case of correlated and anti-correlated data, the growth of the computational time is basically proportional to the growth of dimensionality and cardinality in the same way for all the algorithms. Considering the single graph including the result conducted on both correlated and anti-correlated data, In RSA at the minimum cardinality of 100k the range of response time spans between 5-100s, meanwhile at the maximum cardinality of 1600K spans between 10-500s. JAA graph shows practically the same values than in RSA. Considering instead the growth of dimensionality, represented in [23] through a bar chart, at the minimum of 2 the response time is close to 0, meanwhile at the maximum of 7 the times grows between 140-160s. Two interesting phenomena are worth considering in dimensionality graph: first, the dimensionality grows faster after dimensionality equal to 5, to which correspond a responding time below 60s. The second detail is that at low dimensionalities the difference of response time between JAA and RSA is nonexistent, until again dimensionality becomes equal to 5: after that, RSA increasingly results to be faster than JAA.

### 4.7. Experiments evaluations comparison

Is quite difficult to provide a precise comparison between the complexity of the approaches previously presented for different reasons: first, for every approach are often proposed different algorithms with different complexity growth, which I will try to make a fair average of. A second reason is the different sizes of the data considered, especially in the cardinality, where the variations can be immensely different between approach and approach. Last but not least, every approach has different priorities and characteristics than the others. Despite all this melting pot of dissimilarities, in this section I try to merge such data into an empirical comparison, considering some values for reference. Often in the experiments collected from the papers, a "fixed" cardinality of 100k is taken in account to analyze the complexity, reason why I will also consider it as reference. Also, for the dimensionality, a fixed value of 5 can become another good reference, since included in all the experiments. For that cardinality we can empirically observe that the most efficient approaches are Skyline ordering and R-Skyline, which averagely between the different algorithms requires a computational time included between 0.05-5s the first and between 0.5-5s the second. ORD/ORU approach presents, in my opinion too wide timespan to be easily compared in this way, similarly, even though not that sharply, to K-regret approach where the range of time spans for the order of $10^3$ s. Considering the dimensionality instead, the most efficient for the value "taken" seems to be Skyrank approach and R-Skyline, where the time range spans between 0.3 and 1. ORD/ORU have similar minimum values but some algorithms need up to 100 or 1000 seconds. K-regret instead have such a high diversity of values which, in such general context, the comparison loses meaning since should rather be done considering every single algorithm presented in the bibliography. Skyline ordering, instead, is shown to be the less efficient approach to be considered in this case. UTK seems to have the most stable ratio between the increasing of cardinality/dimensionality and time, reaching peaks of response time close to the average times of the previous six approaches.

## 5. Concluding brief guidelines to choose an approach according to some criteria

Since the criteria for the choice of an approach can be potentially infinite, in this section I will focus mainly on some that can overlap the different problems, solutions and characteristics proposed by the different papers in the bibliography.

If the main goal of the reader is to find a flexible framework with scalability, in this case the suggested one is Skyline Ordering. As introduced in the previous paragraphs, Skyline ordering framework is efficient at providing scalable resolution of arbitrary size constraints on skyline queries. An important drawback that has to be considered though is that, at the variation of the dimensionality, the computational cost of some of the algorithms of this approach drastically increase. The goal of robustness to updates is instead mastered by Skyrank, which makes it the perfect approach to choose if the goal is to deal with skyline points which gets modified often over a dataset. This approach is particularly effective if the size of the dataset considered is vast, and if there is not a complete knowledge about the entire dataset. A certain stability is also provided by K-regret approach, which is insensitive to adding or deleting points that are not optimal for the computation of the output. K-regret though also provide scale invariance against the rescaling of the attribute values, to take in account in the choice if the values in our dataset are known to vary a lot. If a multi objective approach is needed instead, where multiple attributes are required to conduct a choice whose maybe lack completeness or confidence, or if the choice is characterized by complex preferences between them, then the approach suggested is certainly R-Skyline. This approach is mainly efficient when preferences come from a high number of users, like a crowd of people. If the focus instead is the flexibility of personalization, such as dealing with specific preferences of an individual user, then the right approach is ORD/ORU which guarantees a relaxed preference input, allowing flexibility in the specified preferences. This is particularly efficient when the preferences specified by a specific user may change quite easily. If such preferences of the user are instead not certain, but still the focus is to have a user-centric and practical approach, the suggested choice is UTK especially if among the goals to achieve there is also flexibility. Finally, is important to state that results to be efficient even considering voluminous datasets.

# 6. Conclusion

After introducing the main limitations of Top-k queries and skyline queries, I presented 6 new approaches briefly introducing the general definitions, an overview of the algorithms and summarizing experimental results conducted of various datasets. For each of these sections, I wrote a concise comparison between all the different techniques, trying to highlight similarities and dissimilarities. Despite the huge variety of differences, to conclude the paper, I presented a series of short guidelines on how to choose an approach according to some criteria and empirical observations.

# 7. References

[1] Lu H., Jensen C. S. and Zhang Z., "Flexible and Efficient Resolution of Skyline Query Size Constraints", *IEEE TRANSACTIONS ON KNOWLEDGE AND DATA ENGINEERING*, July 2011

[2] Vlachou A. , Vazirgiannis M, "Ranking the sky: Discovering the importance of skyline points through subspace dominance relationships", *Data & Knowledge Engineering*, March 2010

[3] Ciaccia P., Martinenghi D., "Reconciling Skyline and Ranking Queries", *Proceedings of the VLDB Endowment*, August 2017

[4] Nanongkai D, Sarma A. D., Lall A., Lipton R. J., Xu J., "Regret Minimizing Representative Databases", *Georgia Institute of Technology*, 2010

[5] Borzonyi S., Krossmann D., and Stocker K., "The Skyline Operator", *Proc. Int'l Conf. Data Eng. (ICDE)*, 2001

[6] Mouratidis K., Li K., Tang. B., "Marrying Top-k with Skyline Queries: Relaxing the Preference Input while Producing Output of Controllable Size", *SIGMOD '21*, June 2021

[7] Chester S., Thomo A., Venkatesh S., Whitesides S., "Computing kRegret Minimizing Sets", *Proceedings of the VLDB Endowment*, January 2014

[8] Douglas D., Peucker T., "Algorithms for the reduction of the number of points required to represent a digitized line or its caricature", *The Canadian Cartographer*, 1973.

[9] Lofi C., Guntzer U., Balke W., "Efficient computation of trade-off skylines", *EDBT*, 2010

[10] Kaibel V. and M., Pfetsch E., "Some algorithmic problems in polytope theory. In Algebra, Geometry, and software Systems", *TU Berlin*, 2003

[11] Ramer U., "An Iterative procedure for the polygonal approximation of plane curves", *Elsevier Inc*, 1972

[12] Godfrey P., Shipley R., Gryz J., "Algorithms and analyses for maximal vector computation", *VLDB J.*, may 2007

[13] Qian L., Gao J., Jagadish H. V., "Learning user preferences by adaptive pairwise comparison", *PVLDB*, 2015

[14] Berg M. D., Cheong O., Kreveld M. V., Overmars M., "Computational geometry: algorithms and applications", *Springer-Verlag TELOS*, 2008

[15] Chaudhuri S., Dalvi N. N., Kaushik R., "Robust cardinality and cost estimation for skyline operator, Proceedings of International Conference on Data Engineering", *IEEE*, 2006

[16] Fürnkranz J., Hüllermeier E., "Preference Learning", *Springer*, 2010.

[17] Chan C. Y., Jagadish H., Tan K. L, Tung A. K., Zhang Z., "On High Dimensional Skylines," *Proc. Int'l Conf. Extending Database Technology (EDBT)*, 2006.

[18] Papadias D., Tao Y., Fu G., Seeger B., "An Optimal and Progressive Algorithm for Skyline Queries," *Proc. SIGMOD*, 2003.

[19] Chan C. Y., Jagadish H., Tan K. L, Tung A. K., Zhang Z., "Finding K-Dominant Skylines in High Dimensional Space," *Proc. ACM SIGMOD*, 2006

[20] Lin X., Yuan Y., Zhang Q., Zhang Y., "Selecting Stars: The k Most Representative Skyline Operator," *Proc. Int'l Conf. Data Eng. (ICDE)*, 2007.

[21] Koltun V., Papadimitriou C. H., "Approximately Dominating Representatives," *Proc. Int'l Conf. Data Theory (ICDT)*, 2005.

[22] Martinenghi D., Ciaccia P., "Flexible Skylines: Dominance for Arbitrary Sets of Monotone Functions", *ACM Transactions on Database Systems*, December 2020

[23] Mouratidis K., Tang B., "Exact processing of uncertain top-k queries in multi-criteria settings", *Proceedings of the VLDB Endowment*, April 2018

[24] NBA dataset, 2017. www.basketball-reference.com/.